\documentclass[prb,twocolumn,showpacs,superscriptaddress]{revtex4}
\usepackage{epsfig}
\usepackage{amsmath,amssymb}
\usepackage{graphicx}
\usepackage{footnote}

\usepackage{color}

\renewcommand{\omit}[1]{}

\usepackage{units}
\usepackage{extarrows}
\usepackage{float}
\renewcommand{\v}[1]{\ensuremath{\mathbf{#1}}} % for vectors
\newcommand{\avg}[1]{\left< #1 \right>} % for average

\begin{document}

\title{Screened moments and absence of ferromagnetism in FeAl}

\author{A.~Galler,$^{1}$ C.~Taranto,$^{1}$ M.~Wallerberger,$^{1}$ M.~Kaltak,$^{2}$ G.~Kresse,$^{2}$ G.~Sangiovanni,$^{3}$ A.~Toschi,$^{1}$ and K.Held$^{1}$  \\
{\small\em $^1$Institute for Solid State Physics, Vienna University of Technology, 1040 Vienna, Austria \\$^{2}$Center for Computational Materials Science, University of Vienna, 1090 Vienna, Austria\\$^{3}$Institute for Theoretical Physics and Astrophysics, University of W\"urzburg, 97074 W\"urzburg, Germany} 
}

\date{\small\today}

\begin{abstract}
While the stoichiometric intermetallic compound FeAl is found to be paramagnetic in experiment, standard band-theory approaches predict the material to be ferromagnetic. We show that this discrepancy can be overcome by a better treatment of electronic correlations with density functional plus dynamical mean field theory. Our results show no ferromagnetism  down to $\unit[100]{K}$ and since the  
susceptibility is decreasing at the lowest temperatures studied we also do not expect ferromagnetism at even lower temperatures. This behavior is found to originate from temporal quantum fluctuations that screen short-lived local magnetic moments of $\unit[1.6]{\mu_B}$ on Fe.

\end{abstract}
\maketitle

\section{Introduction}

Intermetallic  alloys of iron and aluminum  have a high hardness with a much lower specific weight than steel. Because of this,  their  low costs, and resistance against corrosion and oxidation, FeAl  alloys are often used as lightweight structural materials.
Most puzzling are the  magnetic properties. Here, experiments such as high-field  M\"ossbauer investigations \cite{reissner} indicate no magnetism for stoichiometric FeAl which forms a $B2$ CsCl-type of lattice (two interpenetrating Fe and Al simple  cubic lattices).
Especially the fact that FeAl does not show ferromagnetism in experiment, while  electronic structure calculations within spin-polarized density functional theory (DFT)  predict a ferromagnetic ground state has  drawn  attention to the material: independently of the bandstructure code, DFT orbital basis set and exchange correlation potential  a ferromagnetic ground state with a magnetic moment at the Fe site of about $\unit[0.7]{\mu_B}$ is found. \cite{reissner,deevi,kulikov,sundararajan,chacham}  
Even though  the energy difference between the ferromagnetic and the nonmagnetic state is rather small, the ferromagnetic state is stable over a wide volume range. In fact, only a reduction of the lattice constant by more than 10\% would suppress ferromagnetism. \cite{mohn}  This high stability of the ferromagnetic phase in FeAl suggests that the deviation from experiment is not just a numerical inaccuracy,  but requires a deeper understanding.

Different approaches have been  used hitherto to explain the deviation between spin-polarized DFT and experiment.  One explanation is based on  the fact that the processes used to prepare FeAl often
 "freeze in" chemical disorder. That is, "real" FeAl is usually not fully ordered due to various lattice defects, such as vacancies and antisites, which in turn could have a significant effect on the magnetic properties of the material.
Against this background, there exist several studies concerning the effects of disorder on the magnetic properties of FeAl. \cite{Johnson,Reddy,kulikov,chacham} 
For example, in Ref.\  \onlinecite{Johnson}  the disorder is included  via the coherent potential approximation (CPA) \cite{CPA} in the Korringa, Kohn and Rostoker (KKR) framework, \cite{Korringa,Rostoker} with the  paramagnetic phase described by the disordered local moment approximation (DLM). \cite{DLM}
In agreement with previous DFT calculations, it has been found that ideal FeAl is ferromagnetic. However, even with a small degree of disorder the paramagnetic state, without net magnetization but nonzero local moments, becomes the stable configuration. Thus,  disorder destroys the long-range ferromagnetic order in DFT.
 
However, no ferromagnetism has ever been observed  for stoichiometric ``real'' FeAl, even for samples with very low defect concentration. 
Therefore,  it still remains the question if perfectly ordered FeAl would really be ferromagnetic as predicted by DFT. Indeed, M\"ossbauer experiments \cite{reissner} find magnetic moments only for Fe antistructure atoms (which means Fe atoms sitting on an Al lattice site) and their eight Fe neighbors.

Another possible explanation has been given in Ref.\  \onlinecite{mohn} using
the  DFT+$U$ approach. \cite{anisimov1991} Usually, one would expect DFT+$U$ to yield larger magnetic moments and a stronger tendency towards ferromagnetism than DFT.  For $U$ values ranging from $4$ to $\unit[5]{eV}$ a nonmagnetic state however coexists with the ferromagnetic one in DFT+$U$. 
The ferromagnetic state even disappears for a rather large $U=\unit[5]{eV}$, which offers another explanation of the non-magnetic nature of FeAl.
This rather unusual  DFT+$U$ result can be explained by the changes in the density of states (DOS): increasing  $U$ reduces the DOS at the Fermi level so that according to the Stoner criterion there is no ferromagnetism \cite{mohn} even though the effective exchange is increased by $U$. \cite{lichtenstein} Hence, in a narrow range of $U$,  there is no ferromagnetism in DFT+$U$. \cite{mohn}

In Ref.\ \onlinecite{lichtenstein} it has been argued that this DFT+$U$ result has to be taken with a grain of salt and it has been proposed for the first time that dynamical spin fluctuations suppress ferromagnetism in FeAl. This has been supported by a  dynamical mean field theory (DMFT\cite{Vollhardt,Kotliar}) calculation.\cite{lichtenstein} For $U=\unit[2]{eV}$, FeAl is found\cite{lichtenstein} to be paramagnetic in DFT+DMFT.\cite{Anisimov97a,Lichtenstein98a,Kotliarrev,held} However,  Ref.\ \onlinecite{lichtenstein} only shows a single DFT+DMFT result, the spectral function. The proposed spin fluctuations, the magnetic properties and susceptibility have not been calculated.

Considering these limited  results as well as the improvements of DFT+DMFT in recent years,  a more thorough   analysis is in order. Beyond the first DFT+DMFT spectrum of Ref.\ \onlinecite{lichtenstein}, we study the local and bulk magnetic susceptibility, the magnetic moment and the $k$-resolved spectrum. We also explicitly calculate the local interactions \textit{ab-initio} by constrained random phase approximation (cRPA) and  beyond Ref.\ \onlinecite{lichtenstein} we include the calculated Hund's exchange  in DMFT with its full $SU(2)$ symmetry, since it plays a pivotal role for the magnetic properties. Our results show that while there is
a local moment of even $\unit[1.6]{\mu_B}$  on short time scales, it is screened (suppressed) on longer time scales. This suppression of the local moment 
 occurs on the fs time scale (eV$^{-1}$) and explains why there is eventually no long-range ferromagnetic order.
 
In  Section \ref{Sec:ESCDFT} we present the DFT bandstructure and DOS as well as  the Wannier function projection. Section \ref{Sec:DMFTself} is devoted to the one-particle properties as calculated in DFT+DMFT, i.e., the self-energy as well as the local and $k$-resolved spectral function.  The  DFT+DMFT magnetic properties are discussed in Section \ref{Sec:DMFTmag}, i.e., the local and (zero) ferromagnetic moment as well as the time-dependent local 
susceptibility and bulk susceptibility. Finally, Section \ref{Sec:conclusion} summarizes the results and puts them into context with experiment.

\section{Electronic structure within DFT} \label{Sec:ESCDFT}
\begin{figure}[H]
		\centering
		\includegraphics[width=8cm]{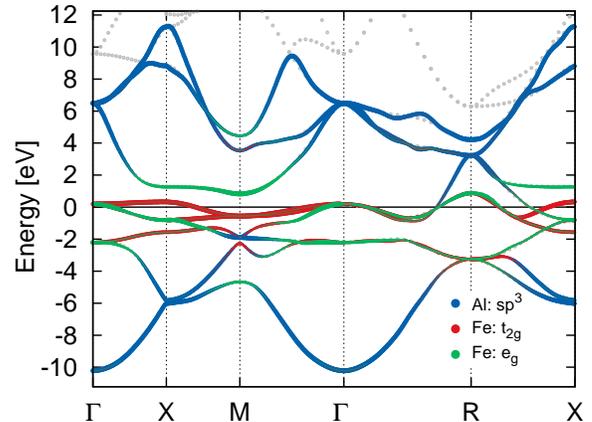}
		\caption[DFT bandstructure ]{Electronic bandstructure of FeAl:
Gray points denote the DFT bandstructure while colored lines show that of
the Wannier projection. The color of the bands indicates the amount of Fe $t_{2g}$
(red), Fe $e_g$ (green)  and Al-$sp^3$ (blue) orbital contribution.}
		\label{fig:bands_vasp}
		\end{figure}

As a first step, we employ the Vienna ab initio Simulation Package (VASP) \cite{hallovasp}
with GGA-PBE functional \cite{PBE1} for calculating the 
bandstructure and density of states of FeAl.
Fig. \ref{fig:bands_vasp} shows the bandstructure of FeAl around the Fermi level. The bands closest to the Fermi level have mainly Fe 3$d$ character 
and are split into  $t_{2g}$ and $e_g$ due to the cubic crystal field.
For these bands we will later include 
electronic correlations by DMFT. However, since the  Fe 3$d$ bands strongly   hybridize with the Al 3$s$ and 3$p$ states, we also include these Al bands (as non-interacting) in our low energy Hamiltonian. 
The corresponding Hamiltonian is obtained by a  projection onto nine maximally localized Wannier orbitals \cite{wannier90}, which reproduce the DFT bandstructure well, 
see Fig. \ref{fig:bands_vasp}.

		\begin{figure}[H]
		\centering
		\includegraphics[width=8.5cm]{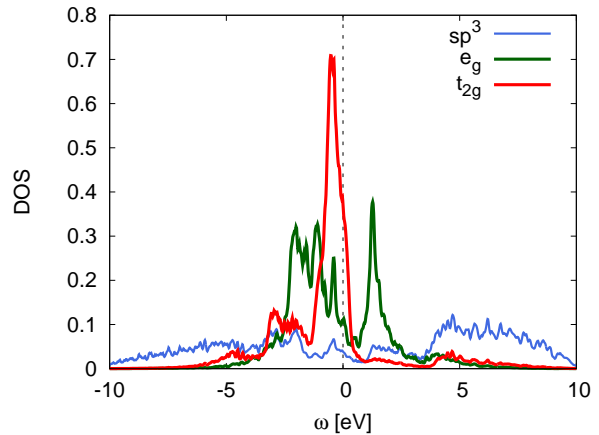}
		\caption[FeAl density of states obtained by DFT]{FeAl density of states obtained by DFT}
		\label{fig:dos_vasp}
		\end{figure}

Fig. \ref{fig:dos_vasp} shows the orbital resolved density of states. It can be seen that the central peak is mainly of $t_{2g}$ character. As the Fe $t_{2g}$ orbitals form only weak bonds with the nearest-neighbor Al atoms, they have a rather small energy dispersion. The Fe $e_g$ states instead  point towards the neighboring Fe atoms and hybridize more strongly. Hence they have a larger bandwidth and split into a bonding- and an antibonding-like part.

\section{DMFT self energy and spectral function} \label{Sec:DMFTself}
After obtaining the low energy Hamiltonian in the basis of Wannier functions, we perform DMFT calculations including the five Fe $d$ orbitals and the 4 Al $sp^3$ orbitals within a so-called $dp$ model.\cite{dp-model} We supplement the DFT-based Wannier Hamiltonian  in DMFT by a local   $d$-$d$ Kanamori interaction, but disregard $d$-$p$ and $p$-$p$ interactions beyond what is already contained in DFT. Note that the hopping terms of the Hamiltonian still contain the full information about the hybridization with the Al $sp^3$ states and charge transfer between $d$ and $sp^3$ orbitals is allowed.

We calculate the screened many-body Coulomb interactions $U$, $U'$ and $J$ by the constrained random phase approximation (cRPA),\cite{RPA1,RPA2} where we  exclude  only the Fe $d$ states from the screening. This is appropriate as  interactions are also applied only to these $d$ states.\cite{silke}    
For our DMFT calculation, we use the average values for the intra-orbital Coulomb interaction $U=\unit[3.36]{eV}$, the inter-orbital Coulomb interaction $U'=\unit[2.36]{eV}$ and the Hund's coupling $J=\unit[0.71]{eV}$. This yields a local,  $SU(2)$-symmetric Kanamori interaction:
 \cite{Kanamori,georges} 
	\begin{eqnarray}\label{Kanamori}
		\hat H_{\rm loc} &=& \sum_mUn_{m\uparrow}n_{m\downarrow} \nonumber \\
			&+& \sum_{m\neq m',\sigma}[U'n_{m\sigma}n_{m',-\sigma}+(U'-J)n_{m\sigma}n_{m'\sigma}]\nonumber\\
			&+& \sum_{m\neq m'}Jc^\dagger_{m\uparrow}c^\dagger_{m'\downarrow}c_{m\downarrow}c_{m'\uparrow}\nonumber\\
			&+& \sum_{m\neq m'}Jc^\dagger_{m\uparrow}c^\dagger_{m\downarrow}c_{m'\downarrow}c_{m'\uparrow} .
	\end{eqnarray}
Here,  $c^\dagger_{m\sigma}$ ($c_{m\sigma}$) creates (annihilates) an electron
with spin $\sigma$ in  the Fe $3d$ orbital $m$;  $n_{m\sigma}=c^\dagger_{m\sigma}c_{m\sigma}$. We  employ the double counting correction of the fully localized limit, \cite{anisimov1993} and validate that a  difference of  $\unit[2.5]{eV}$ in the double counting  does not change our findings (not shown).

For the solution of the DMFT impurity problem we use a continuous-time quantum Monte Carlo (CT-QMC) algorithm in its hybridization expansion (CT-HYB) in the version of Ref. \onlinecite{nico}, for a review see Ref.\ \onlinecite{Gull}.
Especially with regard to the magnetic properties that we will compute, it is important to employ the rotationally invariant form of the interaction term $H_{\rm loc}$ above, including a pair-hopping and a spin-flip term, and not only density-density contributions. As for the CT-HYB, we note that it is essential 
to truncate the outer states  for the evaluation of the local fermionic trace only at high energies, especially at high temperatures.

\begin{figure}[H]
		\centering
		\includegraphics[width=8.5cm]{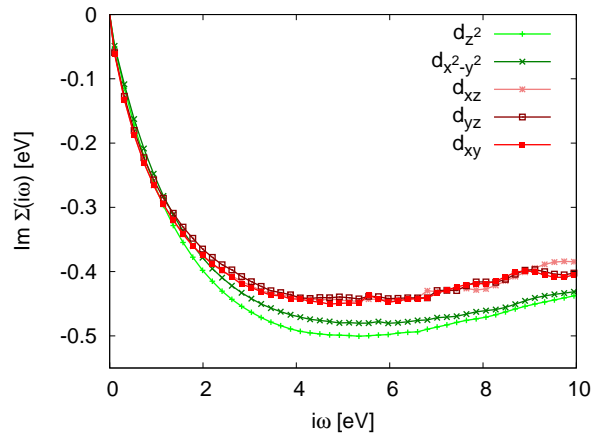}
		\caption[DMFT self energies for the Fe $d$ orbitals]{DMFT self energies for the Fe $d$ orbitals (at inverse temperature $\beta=\unit[30]{eV^{-1}}$ corresponding to $390\,$K). The extracted quasiparticle weight is $Z=\unit[0.75]{}$.}
		\label{fig:siw}
		\end{figure}

Fig. \ref{fig:siw} shows the imaginary part of the DMFT self energy $\Sigma(i\omega)$ on the Matsubara axis for all five Fe $d$ orbitals. In order to avoid all uncertainties related to an analytical continuation, we calculate the quasiparticle weight $Z$ directly from the self energy on the Matsubara axis $Z=1/(1-\Im(\partial\Sigma(i\omega)/\partial (i\omega))|_{i\omega\rightarrow0})$. This yields a value of $Z=\unit[0.75]{}$,  essentially the same for all $3d$ orbitals. This $Z$ value would indicate a rather weakly correlated material.

\begin{figure}[H]
		\centering
		\includegraphics[width=8.5cm]{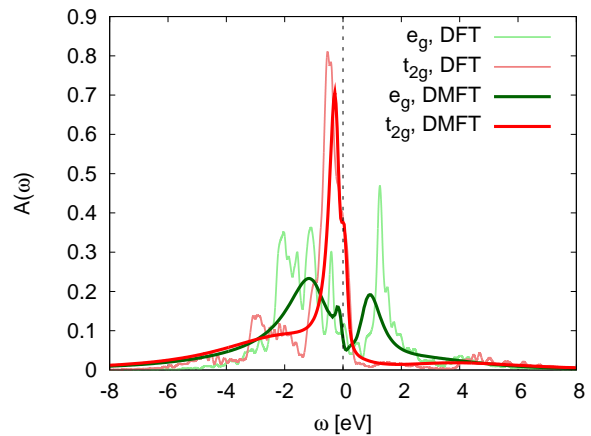}
		\caption[]{DMFT spectral function for the  Fe $d$ orbitals at $\beta=\unit[30]{eV^{-1}}$, compared to DFT ($\omega=0$ corresponds to the Fermi level).} 
		\label{fig:maxent}
		\end{figure}

The corresponding spectral function $A(\v{k},\omega)=-1/\pi{ }\Im(G(\v{k},\omega+i0^+))$ is shown in Fig.\ref{fig:maxent} on the real frequency axis, for which  an analytic continuation using a stochastic version of the maximum entropy method has been used.\cite{maxent}
In comparison to the DFT DOS, both occupied and empty states are slightly shifted towards the Fermi energy due to the Fermi-liquid renormalization. There is no evidence of pronounced upper and lower Hubbard bands and one can only observe a weak increase of the spectral weight at high frequencies.
In agreement with  Ref.\ \onlinecite{lichtenstein}, we find that the spectral function at the Fermi level is essentially the same in DFT+DMFT as in DFT. We did not perform charge-self-consistent calculations since the difference in the occupation of the $d$-orbitals between the DFT-derived Hamiltonian and DMFT is very small. In DFT, we have $\unit[4.8]{}$ electrons in the $t_{2g}$ and $\unit[2.5]{}$ electrons in the $e_g$ states out of 11 electrons per unit cell, in DMFT the $t_{2g}$ orbitals are occupied with $\unit[4.8]{}$ and the $e_g$ orbitals with $\unit[2.6]{}$ electrons.  Also the changes in the one-particle spectrum are rather small. Note, only if DMFT  alters the spatial charge distribution $\rho({\mathbf r})$ considerably, charge self-consistency would have an effect. Thus, we expect changes by charge-self-consistency to be small.

\begin{figure}[H]
		\centering
		\includegraphics[width=8.5cm]{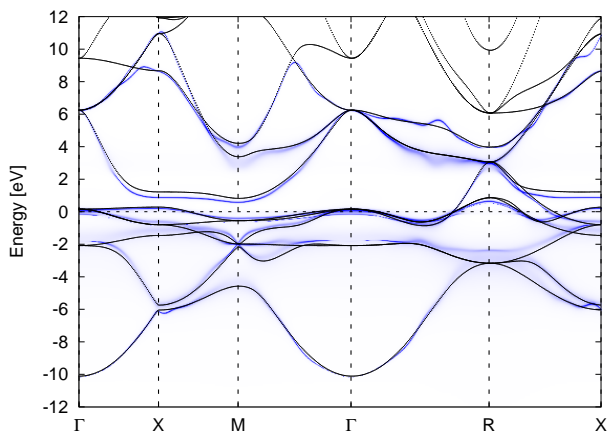}
		\caption[]{ DMFT  $k$-resolved spectral function (blue) compared to the DFT bandstructure (black).}
		\label{fig:bands_dmft}
		\end{figure}

Fig. \ref{fig:bands_dmft} presents the corresponding $k$-resolved spectrum which shows that also the DFT+DMFT bands essentially follow the DFT bandstructure. The most noteworthy effects are again a slight shift towards the Fermi level, i.e., a quasiparticle renormalization and a broadening of the bands, especially of the $d$ bands located around the Fermi level. Hence, regarding only  single-particle  quantities, FeAl seems to exhibit only weak correlation effects. However, this picture changes when considering also two-particle quantities, namely the magnetic susceptibility.

\section{DMFT magnetic properties} \label{Sec:DMFTmag}
In order to study the magnetic properties of FeAl within DFT+DMFT, we compute the local magnetic susceptibility, represented by the two-particle spin-spin correlation function

\begin{eqnarray}\label{chi_loc}
	\chi_{\rm loc}(\tau)=\sum_{m,n}\chi_{\rm loc}^{m,n}(\tau)=g^2\sum_{m,n}\avg{S_z^m(\tau)S_z^n(0)}
\end{eqnarray}
	
with $m$ and $n$ being the orbital indices of the five Fe $d$ orbitals, $\tau$ the imaginary time, and $g\approx2$ the gyromagnetic factor for the electronic spin. $S_z^m(\tau)=1/2(n_{m \uparrow}(\tau)-n_{m \downarrow}(\tau))$ is the $z$-component of the spin operator of orbital $m$, expressed in terms of the corresponding density operators $n_{m \sigma}=c_{m \sigma}^{\dagger}c_{m \sigma}^{\phantom{\dagger}}$.

\begin{figure}[H]
		\centering
		\includegraphics[width=8.5cm]{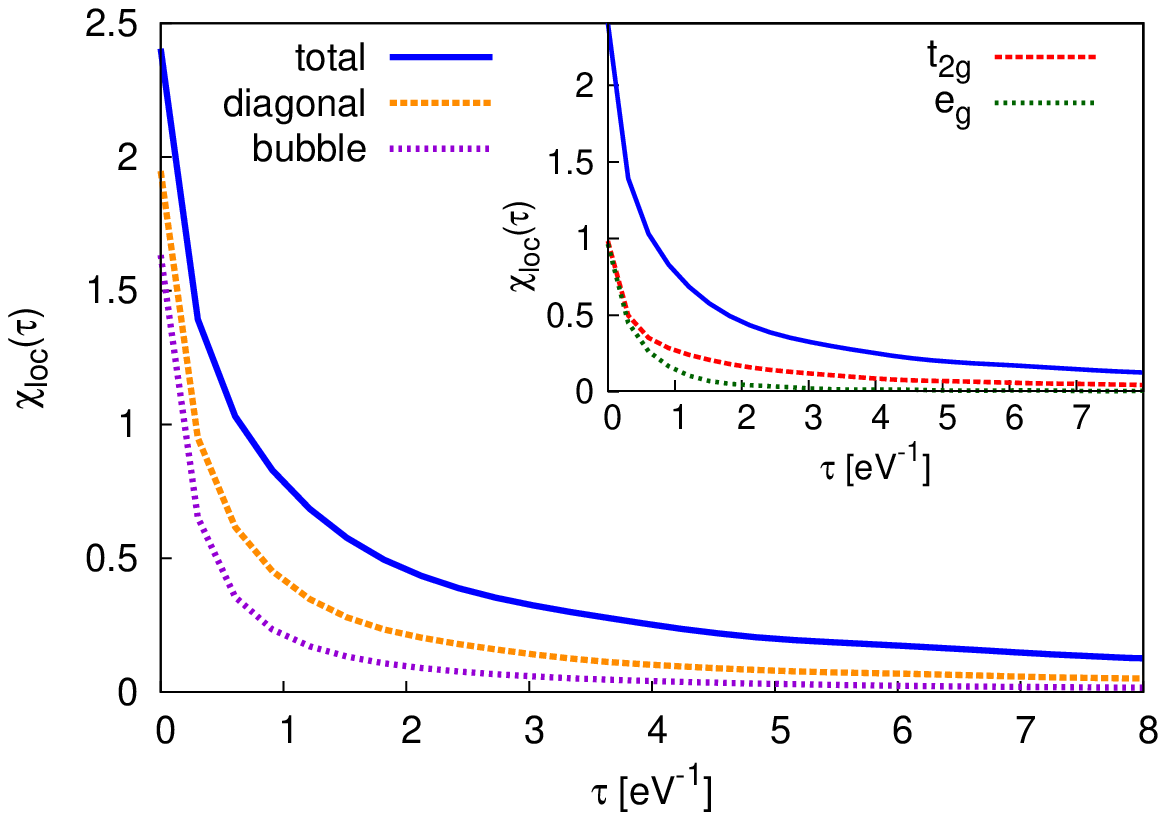}
		\caption[]{Local magnetic susceptibility  $\chi_{\rm loc}$ as a function of (imaginary) time $\tau$  for $\beta=\unit[30]{eV^{-1}}$. Besides the total susceptibility, also its diagonal and bare-bubble contribution are shown, as well as (in the inset) its $e_g$ and $t_{2g}$ contribution.\footnote{The results shown in Fig. \ref{fig:chi_loc} have been checked to be stable over a temperature range from $\beta=\unit[10]{eV^{-1}}$ to $\beta=\unit[35]{eV^{-1}}$. }}
		\label{fig:chi_loc}
		\end{figure}

Technically speaking, $\chi_{\rm loc}(\tau)$ is obtained by first measuring the generalized magnetic susceptibility $\chi_{\rm loc}(i\nu,i\nu',i\omega)$ of the converged DMFT impurity model by means of CT-HYB quantum Monte Carlo sampling. Thereby, $\chi_{\rm loc}(i\nu,i\nu',i\omega)$ automatically contains all vertex corrections to the bare (DMFT) bubble spin susceptibility. The sum over the fermionic Matsubara frequencies $\nu$ and $\nu'$ and a Fourier transform, $\chi_{\rm loc}(\tau)=1/\beta\sum_{i\omega}e^{-i\omega\tau}\chi(i\omega)$, finally lead  to $\chi_{\rm loc}(\tau)$.
Here, for  the large frequency asymptotics, the bare bubble contribution, Eq.\ (\ref{Eq:3}), which is known  on a larger frequency grid and an additional fitting function of the form $1/\nu^2$ have been used. The results for $\chi_{\rm loc}(\tau)$  are shown in Fig. \ref{fig:chi_loc} for $\beta=\unit[30]{eV^{-1}}$ (for lower temperatures the numerically feasible frequency box becomes too small).

The solid, blue curve in Fig. \ref{fig:chi_loc} corresponds to the total magnetic susceptibility $\chi_{\rm loc}(\tau)$ of Eq. (\ref{chi_loc}). The dashed, orange curve instead represents the orbital-diagonal contribution $\sum_m \chi_{\rm loc}^{m,m}(\tau)$. The dotted, purple curve is the bare-bubble contribution $\chi_{\rm loc}^0(\tau)$, which neglects vertex corrections and is obtained by directly convoluting the DMFT Green functions $G_{m}(i\nu)$:

\begin{eqnarray}
	\chi_{\rm loc}^0(i\omega)=-\frac{1}{\beta}\sum_{i\nu,m\sigma}G_{m\sigma}(i\nu)G_{m\sigma}(i\nu+i\omega)\label{Eq:3}
\end{eqnarray}

The significant difference between the bare-bubble contribution and the susceptibility including vertex corrections in Fig. \ref{fig:chi_loc} reveals that
electronic correlations actually play a major role in FeAl, more than it  could be expected  from the weak quasiparticle renormalization. Fig. \ref{fig:chi_loc} also shows that the enhancement of $\chi_{\rm loc}(\tau)$ stems approximately in equal parts from an enhancement of the intra-orbital contribution (diagonal part) and additional inter-orbital (off-diagonal) contributions, which are not present in the bare-bubble susceptibility. 

The local susceptibilities in Fig. \ref{fig:chi_loc} show a rather fast and strong decay in $\tau$. Here, the value of $\chi_{\rm loc}(\tau)$ at $\tau=0$ can be interpreted in terms of  the instantaneous, local magnetic moment. The observed decay in $\tau$ reflects a dynamical screening of this local magnetic moment due to quantum fluctuations.
 Thus, we can conclude that dynamical quantum fluctuations significantly reduce the local magnetic moment in FeAl.

Fitting $\chi_{\rm loc}(\tau)$ to an exponential between $\tau=\unit[0]{eV^{-1}}$ and $\tau=\unit[5]{eV^{-1}}$ yields a time scale for the screening of $\tau_s=\unit[1.03]{eV^{-1}}=\unit[4.02]{fs}$. The inverse of $\tau_s$ is the energy scale associated with the screening which is essentially the bare bandwidth if we have a noninteracting system, the width of the central peak if we consider the interacting bubble, and the Kondo temperature for the interacting system with vertex corrections. This Kondo temperature is smaller than the  width of the central peak.\cite{kondo} Hence the decay with vertex corrections should be slower. Indeed, in Fig. \ref{fig:chi_loc}  the total $\chi_{\rm loc}(\tau)$ decays slower than the bubble contribution.  For a related analysis, how to interpret the susceptibility as a function of imaginary time and how the local, fluctuating magnetic moment reflects as a pronounced low-energy peak in the local neutron spectra, see \onlinecite{iron-pnictides,Prokopjev91,iron-pnictides2}.

 In the inset of  Fig.\ \ref{fig:chi_loc}, we separate the $e_g$ and $t_{2g}$ contributions of the susceptibility. These two contributions are rather independent 
as one clearly sees from the longer time scale on which the $t_{2g}$
susceptibility decays. This different decay rate can be explained by the considerably more narrow  $t_{2g}$  bandwidth and hence stronger correlations of the $t_{2g}$ orbitals.  
If Hund's exchange was the major player, on  the other hand, one would expect a stronger coupling of $e_g$ and $t_{2g}$ susceptibility, and a decay on a similar time scale.

This all suggests that the Hund's rule exchange $J$, which mainly drives the inter-orbital contribution, is not exceedingly important in FeAl. This is 
in contrast to other Fe-based compounds such as  the iron-based superconductor LaFeAsO.\cite{silke,iron-pnictides}  which have been classified as  Hund's metals. \cite{georges,Haule09}

\begin{figure}[H]
		\centering
		\includegraphics[width=8.5cm]{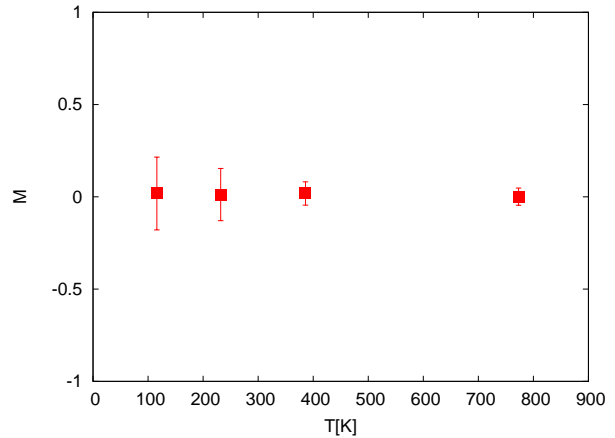}
		\caption[]{DFT+DMFT magnetization for different temperatures. The ferromagnetic moment is zero within the error bars, FeAl is a paramagnet.}
		\label{fig:magn}
		\end{figure}

From the  local magnetic properties, we now turn to the bulk magnetic susceptibility and the  long-range ordered ferromagnetic moment. 
 Fig. \ref{fig:magn} shows the ordered magnetic moment, which
has been obtained by breaking the spin symmetry in the first DMFT iteration
so that 
 the system can either stabilize a para- or ferromagnetic solution.
 As Fig. \ref{fig:magn} clearly shows, the ordered ferromagnetic moment is 
zero down to a temperature of $\unit[100]{K}$. Thus, in the investigated temperature range, FeAl is paramagnetic in DFT+DMFT, in agreement with experiment but in contrast to DFT.

This result is also supported by the calculation of the bulk ferromagnetic susceptibility in DFT+DMFT. To this end, we have applied a small magnetic field of  $H=\unit[0.005]{eV}$, checked (for some temperatures) that this is still in the linear $M$ vs.\ $H$ regime (which further confirms the paramagnetic phase) and calculated $\chi(\vec{q}=0)=M/H$ at this $H$. This way all vertex corrections are included; and this quantity allows to determine whether there is a second order phase transition towards a ferromagnetic phase or not. Prospectively competing phases with a different wave vector $\vec{q}$ are however not accessible this way.

\begin{figure}[H]
		\centering
		\includegraphics[width=8.5cm]{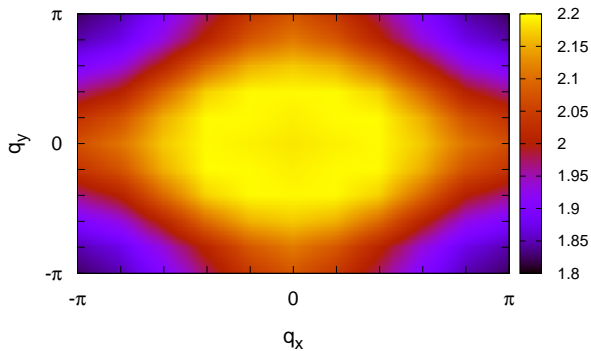}
		\caption[]{Susceptibility $\chi^0(\vec{q},i\omega=0)$ vs. $q_x$ and $q_y$ (at $q_z=0$) calculated from the DMFT $d$-electron Green functions at $\beta=\unit[100]{eV^{-1}}$. The maximum at $\vec{q}=0$ indicates that  without vertex corrections ferromagnetism is the leading instability.}
		\label{fig:chi0_q}
		\end{figure}

The  full $\vec{ q}$-dependent susceptibility $\chi(\vec{q},i\omega)$ could in principle be obtained  by solving the Bethe-Salpeter equation. Unfortunately, this is computationally too demanding for five orbitals at low temperatures. For the same reason the local  susceptibility $\chi_{\rm loc}(\tau)$ could only be calculated reliably down to  $\beta=\unit[30]{eV^{-1}}$.
But to gain at least some insight whether ferromagnetism or magnetic phases with other $\vec{q}$-vectors prevail,
we study the bare bubble susceptibility
$\chi^0(\vec{q},i\omega=0)=-\frac{1}{\beta}\frac{1}{N_k}\sum_{i\nu,\vec{k},m,n,\sigma}G_{mn\sigma}(i\nu,\vec{k})G_{nm\sigma}(i\nu,\vec{k}+\vec{q})$, which does not include vertex corrections. The result
shown in Fig. \ref{fig:chi0_q} indicates that $\vec{q}=0$ is the leading instability.
\footnote{While the Stoner criterion $I \chi(q=0) > 1$ would predict ferromagnetism for $I=U$ or $I=J$, it is known that this criterion largely overestimates the tendency towards ferromagnetism.\cite{ferro1, ferro2}} Thus, in the following we will focus on $\chi(\vec{q}=0)$.

\begin{figure}[H]
		\centering
		\includegraphics[width=8.5cm]{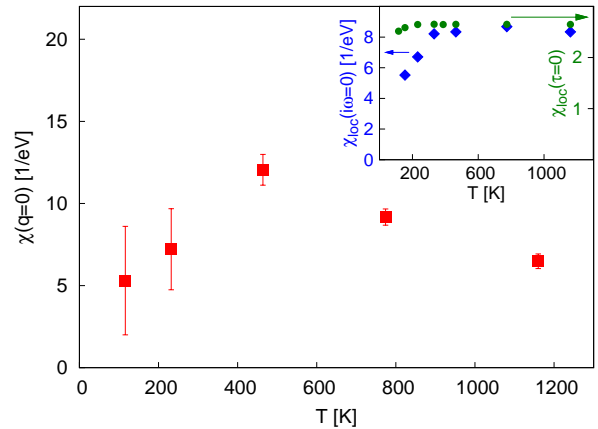}
		\caption[]{Ferromagnetic susceptibility $\chi(q=0)$ as a function of temperature. The inset shows the temperature dependence of the corresponding local quantities: the local magnetic susceptibility ($\chi_{\rm loc}(i\omega=0)$, diamonds) and the instantaneous one ( $\chi_{\rm loc}(\tau=0)$, circles), from which the local magnetic moment can be estimated ($\propto \sqrt{\chi_{\rm loc}}$).}
		\label{fig:chi_q0}
		\end{figure}

The temperature dependence of the susceptibility $\chi(\vec{q}=0)=M/H$ including vertex corrections is shown in Fig. \ref{fig:chi_q0}. Upon decreasing temperature, we first notice an increase of the susceptibility.  However, below $400$K, the susceptibility decreases again.
This clear trend of a {\sl reduction} of the susceptibility by decreasing $T$ makes the onset of a ferromagnetic order at lower temperatures extremely unlikely. We note that a marked low-$T$ reduction of $\chi(\vec{q}=0)$ has been also reported experimentally\cite{Klingeler2010} and theoretically\cite{Skornya2011} in the iron-pnictide compound LaFaAsO. There, this behavior of $\chi(\vec{q}=0)$ coexists with an opposite (increasing) trend of the local magnetic susceptibility $\chi_{\rm loc}$. \cite{Skornya2011}  Hence,  the unusual low-$T$ reduction of $\chi(\vec{q}=0)$  has been attributed to specific features of the one-particle spectral function of LaFeAsO, displaying significant temperature variations near the Fermi level. By performing the same analysis for FeAl we find, however, that the low-$T$ behavior of $\chi_{\rm loc}$  (inset of Fig. \ref{fig:chi_q0}) and  $\chi(\vec{q}=0)$ (main panel) is qualitatively very similar: both show a visible reduction for $T < 400$K. In the very same temperature interval, a slight reduction of the instantaneous local moment ($\chi_{\rm loc}(\tau =0)$, inset) is also found, which is a typical behavior in the Fermi liquid regime, as described by the DMFT.
 
Hence, in FeAl, the role played by emerging low-energy structures of the spectral function appears to be less important than in LaFeAsO. Rather, the trend of $\chi(\vec{q}=0)$ in FeAl may simply reflect the corresponding low-$T$ reduction of the local magnetic moment ($\propto \sqrt{\chi_{\rm loc}}$), in particular of the screened one. The latter can be ascribed to the enhanced metallic coherence of the low-temperature region, which is a general effect of local correlations in the Fermi-liquid regime.

\section{Conclusion}\label{Sec:conclusion}
In the present work, we studied FeAl by means of DFT+DMFT. In particular, we investigated the effects of electronic correlations on the magnetic properties since standard spin-polarized DFT calculations yield a ferromagnetic ground state whereas in experiment it is a paramagnet. From DFT, we constructed a 9-band Wannier Hamiltonian  with four Al $sp^3$ orbitals and  five Fe $d$ orbitals. 
For the latter we include a local $SU(2)$-symmetric Kanamori interaction with $U=\unit[3.36]{eV}$, $U'=\unit[2.36]{eV}$ and $J=\unit[0.71]{eV}$ as obtained from cRPA and solve the many-body problem by DMFT(CT-QMC).

On the one particle level, the self energy and Green function suggest rather weak electronic correlations with a quasiparticle renormalization of only $Z=0.75$ and no evidence of pronounced upper and lower Hubbard bands. In DFT+DMFT
we calculate from the spin-spin correlation function an equal-time local magnetic moment of  $\unit[1.6]{\mu_B}$ which is twice as large as the magnetic moment in spin-polarized DFT. It is also much larger than the bubble contribution which demonstrates that electronic correlations are pivotal for the two-particle quantities in FeAl. 

Even more importantly,  our results show that the moment is fluctuating in time and screened on the fs time scale. This explains why we also do not find  long-range ferromagnetic order. According to our DFT+DMFT study, FeAl is paramagnetic
with a maximum in the ferromagnetic susceptibility around room temperature and no tendency towards long range magnetic order in the temperature range studied.

Previously, it has been proposed that disorder and a spin-glass behavior might explain the missing ferromagnetic moment in experiment. Our results show that if temporal fluctuations are taken into account the moment is actually screened on short time scales. There is hence, even for a perfect lattice, neither  ferromagnetism nor a local moment on longer time scales. If the magnetic moments were constant in time and  spatially disordered, M\"ossbauer experiments, which probe the {\em local} magnetic moment, are  in principle able to reveal it. The fluctuating local moment  on the fs time scales, can however not be observed   in  M\"ossbauer spectroscopy which cannot resolve such short time scales. Hence, the   M\"ossbauer experiments, \cite{reissner} which show no local magnetic moment for stoichiometric  FeAl, seem to better agree with a magnetic moment fluctuating in time as we find in DFT+DMFT than with a moment fluctuating in space.

\section*{Acknowledgments}
We are grateful to P. Mohn, S. Khmelevskyi, R. Podlucky, M. Reissner, M. Karolak and J. Tomczak for valuable discussions.
 We acknowledge financial support  by the Austrian Science Fund (FWF)  through Doctoral School  Solids4Fun W1243 (AG), SFB ViCoM F41
(MW,GK,AT),  DFG research unit FOR 1346 (MK,CT,GS), SFB 1170 ToCoTronics and by the European Research Council under the European Union's Seventh Framework Program (FP/2007-2013)/ERC through grant agreement n.\ 306447 (AG,KH).
 The computational results presented have been obtained using the Vienna Scientific Cluster (VSC).

\vfill\eject

\end{document}